\documentclass[12pt,a4paper]{article}
\usepackage{epsfig,amssymb,amsmath,psfrag}

\def \bl  {\begin{align*}}
\def \el  {\end{align*}}

\def \be  {\begin{equation}}
\def \ee  {\end{equation}}
\def \ba  {\begin{eqnarray}}
\def \ea  {\end{eqnarray}}
\def \baa {\begin{eqnarray*}}
\def \eaa {\end{eqnarray*}}
\def \bb  {\begin {thebibliography} }
\def \eb  {\end{thebibliography}}
\def \lab #1 {\label{#1}}

\def \qqquad {\qquad\quad}

\newcommand{\beq}{\begin{equation}}
\newcommand{\eeq}{\end{equation}}
\newcommand{\beqa}{\begin{eqnarray}}
\newcommand{\eeqa}{\end{eqnarray}}

\newcommand{\tlam}{\tilde{\lambda}}

\def \Tr {\mathop{\rm Tr}\nolimits}

\def \e  {\mathop{\rm e}\nolimits}

\renewcommand{\a}{\alpha}
\newcommand{\adt}{{\dot{\alpha}}}
\newcommand{\bdt}{{\dot{\beta}}}
\renewcommand{\b}{\beta}

\def\e{\epsilon}
\def\d{\delta}
\def\lam{\lambda}
\def\tlam{\tilde{\lambda}}
\def\th{\theta}
\def\cN{\mathcal{N}}
\def\cP{\mathcal{P}}
\def\cW{\mathcal{W}}
\def\cZ{\mathcal{Z}}
\def\AA{\mathcal{A}}
\def\BB{\mathcal{B}}
\def\CC{\mathcal{C}}

\def\MM{\mathcal{M}}

\def\NN{\mathcal{N}}

\def\gg{\mathfrak{g}}

\def\del{\partial}

\def\l<{\langle}
\def\r>{\rangle}

\def\XXint#1#2#3{{\setbox0=\hbox{$#1{#2#3}{\int}$}
     \vcenter{\hbox{$#2#3$}}\kern-.5\wd0}}

\renewcommand{\title}[1]{\vbox{\center\LARGE{#1}}\vspace{5mm}}
\renewcommand{\author}[1]{\vbox{\center#1}\vspace{5mm}}

\begin{document}

\thispagestyle{empty}

\begin{flushright}
HU-EP-11/34\\
~
\end{flushright}

\vskip2.2truecm
\begin{center}
\vskip 0.2truecm {\Large\bf
{\Large Yangian Symmetry in $\cN = 4$ super Yang-Mills}
}\\
\vskip 1truecm
{\bf  
}

\vskip 0.4truecm

\begingroup\bf\large
L. Ferro
\endgroup
\vspace{0mm}

\begingroup
\textit{Institut f\"ur Physik, Humboldt-Universit\"at zu Berlin, \\
Newtonstra{\ss}e 15, D-12489 Berlin, Germany}\par
\texttt{ferro@physik.hu-berlin.de\phantom{\ldots}}
\endgroup

\end{center}

\vskip 1truecm 
\centerline{\bf Abstract} 

In this paper we review recent results on symmetries in $\cN = 4$ super Yang-Mills theory. Symmetries are of invaluable help in studying and constraining the scattering amplitudes, and there has been a lot of progress in recent years concerning this topic. It has been realised that the ordinary superconformal symmetry is not the full symmetry of this theory. There is indeed a dual superconformal symmetry, and together they form a Yangian structure, pointing to the underlying integrability of the theory.
Here we give an overview of the Yangian algebra and of its action  on the scattering amplitudes, at tree and loop level. This article is an invited review for the Focus Issue \emph{Gauge/String Duality 2011} of Advances in High Energy Physics.

\medskip

 \noindent

\newpage
\setcounter{page}{1}\setcounter{footnote}{0}

\section{Introduction}

Much attention has been recently devoted to $\NN = 4$ supersymmetric Yang-Mills  theory (SYM). It reveals indeed many interesting features, which turn out to be very useful even for less supersymmetric theories.  
In particular, it has been shown that in the planar limit it possesses a very large symmetry. In addition to the ordinary superconformal, indeed, it has been discovered the presence of a distinct hidden symmetry, not visible at the Lagrangian level. This dual superconformal symmetry  was revealed at weak coupling by considering the amplitudes in a dual space and it  is related to the duality between  maximally-helicity violating (MHV) amplitudes and light-like Wilson loops.   
The combination of the two symmetries forms a Yangian structure, which points once more at some underlying integrability of $\NN = 4$ SYM.
The exactness of the Yangian symmetry at the level of the  amplitudes is  obscured by the infrared divergences, which arise when actually computing the scattering amplitudes.  Indeed, even if  tree-level amplitudes are invariant under the symmetry (except on singular kinematic configurations), at loop level  the divergences break it. Much effort has been put into the understanding of its fate at higher loop orders.
Indeed the importance of symmetries in a theory is remarkable, as they strongly constrain the structure of the scattering amplitudes, acting in a predictive way. For instance, the anomalous Ward identity of the dual conformal symmetry indicated that the Bern-Dixon-Smirnov ansatz for the finite part of the $n$-point MHV scattering amplitude could fail starting from six points.
Therefore, the analysis of the symmetry properties of a theory is fundamental in understanding its behaviour.

This paper aims to present an overview of these topics, with particular emphasis on the Yangian symmetry. 
 In the next section we review basic notions about Yangian algebras and representations, with special attention to the oscillator representation. Then in section \ref{ampl}, we give some details on the structure of superamplitudes and on the superconformal and dual superconformal symmetry of $\cN = 4$ SYM. We show how they combine in a Yangian structure and how this acts on the amplitudes.  We also discuss the duality between Wilson loops and MHV  amplitudes.
In section \ref{Yinvariants} we deal with the topic of Yangian invariants. In particular, we begin by  reviewing the Grassmannian formulas, which compute leading singularities through contour integrals. Then in the very last part we discuss  a recently proposed finite function, constructed from a combination of Wilson loops, which is Yangian invariant at one loop at least in a special kinematics.


\section{Yangian algebra and representations}
\label{Yalgebra}
 
We begin our overview by recalling some major notions about Yangian algebras, which will be useful later on. We focus on the non-graded case (Lie algebras), but everything can be extended straightforwardly to superalgebras. Detailed introductions to this subject can be found for instance in \cite{Chari:1991xx,Bernard:1992ya,MacKay:2004tc}.

The definition of Yangian algebra was first introduced by Drinfeld in \cite{Drinfeld:1985rx,Drinfeld:1986in}. 
Let us call  $\gg$ the simple Lie algebra spanned by the generators $J_a$:
\beq
\label{levelzero}
[J_a, J_b] = f_{ab}^{~~c} J_c \,,
\eeq
where $f_{ab}^{~~c}$ are the structure constants of $\gg$ and{\footnote{In the following we will use lower-case letters $a$ for the adjoint representation and upper-case $A$ for the fundamental one.}} $a = 1,\ldots, \mathrm{dim \gg} $. The Lie algebra is an Hopf subalgebra of the Yangian wih trivial coproduct $\Delta : \gg \rightarrow \gg \otimes \gg$, which acts on the generators in the following way
 \begin{align}
 \label{liecopr}
\Delta (J_a) &= J_a \otimes 1 + 1 \otimes J_a \,.
\end{align}
The Yangian $Y(\gg)$ of a Lie algebra $\gg$ is then the Hopf algebra generated by the set of $J_a$'s (which form the so-called level zero) together with another set $J^{(1)}_a$, the level one,  which obeys
\beq
\label{levelone}
[J_a, J^{(1)}_b] = f_{ab}^{~~c} J^{(1)}_c \,,
\eeq
therefore transforming in the adjoint representation of $\gg$. The coproduct $\Delta : Y(\gg) \rightarrow Y(\gg) \otimes Y(\gg)$ acts  non-trivially on the level-one generators
 \begin{align}
 \label{coproduct}
\Delta (J^{(1)}_a) &= J^{(1)}_a \otimes 1 + 1 \otimes J^{(1)}_a +  \gamma f_{a}{}^{cb} J_b \otimes J_c\,,
\end{align}
where $\gamma$ is any number different from zero\footnote{Since it does not change the relations below, we will choose it to be one.}.
By requiring $\Delta (J^{(1)}_a) $ to be an homomorphism,  the commutator between level-one generators is constrained by  the Serre relations:
\begin{align}
\label{serre}
&[J^{(1)}_a , [J^{(1)}_b,J_c]] +  [J^{(1)}_b,[J^{(1)}_c,J_a]] + [J^{(1)}_c,[J^{(1)}_a,J_b]] \notag \\
&= \{J_l,J_m,J_n\} f_{ar}{}^{l} f_{bs}{}^{m} f_{ct}{}^{n} f^{rst} \,,
\end{align}
where $\left\{A,B,C\right\}$ is the symmetrized product of the three generators $A$, $B$ and $C$.
Given these relations, the infinite number of levels of the Yangian algebra is completely generated. For instance the second level will be constructed from the commutation relations of level-one generators: 
 \beq
[J^{(1)}_a, J^{(1)}_b] = f_{ab}^{~~c} J^{(2)}_c + h(J, J^{(1)})\,
\eeq
with $h(J, J^{(1)})$ a function of the level zero and one being constrained by (\ref{serre}). The coproduct helps to build general representations from simple ones. For instance, if the space where we act is a multi-site space, given by the tensor product representation of $n$ vector spaces, then the generators will be defined as follows:
\begin{align}
\label{sumzero}
J_a &= \sum_{i=1}^n J_{a,i} \,,\\
J^{(1)}_a &= \sum_{1\leqslant i < j \leqslant n} f_{a}^{~cb} J_{b,i} J_{c,j} + \sum_{k=1}^n c_k J_{a,k} \,,
\label{bilocal}
\end{align}
where the definition (\ref{bilocal}) of level-one generators in terms of the level-zero set is the so-called bilocal formula and it is given by the coproduct (\ref{coproduct}). We will show explicitly how the coproduct actually works in the next section.
The $c_k$ which enter in the bilocal formula are free parameters which will show up again in the last section.

For the reader familiar with loop algebras, we note that the Yangian is somehow a "half-loop" algebra, being defined by
\beq
[t_{a}^{(n)}, t_{b}^{(m)}] = f_{ab}^{~~c} t_{c}^{(n+m)} \,,
\eeq
where $n, m \in \mathbb{Z}^{+} $ define the level of the generators.

Another presentation of $Y(\gg)$ is given by the R-matrix. Here we just want to mention that the generators of the Yangian algebra can be arranged in a generating function 
\begin{equation}
t^{AB}(u) = \sum_{n=0}^{\infty} u^{-n} t_n^{AB} 
\end{equation}
where $u$ is the spectral parameter, $n$ is the level and $t_0^{AB} = \delta^{AB}$. If we consider $Y(gl(N))$, by introducing an auxiliary space $E_{AB}$, one can define a matrix $T(u)$
\begin{equation}
T(u) = \sum_{A,B=1}^N t^{AB}(u) \otimes E_{AB} \,
\end{equation}
and a matrix $R_{12}(u-v)$
\begin{equation}
\label{Rmatrix}
R_{12}(u-v) = 1_N \otimes 1_N - \frac{1}{u-v} P_{12}
\end{equation}
where $P_{12}$ is the permutation operator $
P_{12} =  \sum_{A,B=1}^N E_{AB} \otimes E_{BA} \,.
$
Then the so-called RTT relation
\begin{equation}
\label{rtt}
R_{12}(u-v) T_1(u) T_2(v) = T_2(v) T_1(u) R_{12}(u-v)
\end{equation}
is equivalent to all the Yangian relations when expanded in  series of $\frac{1}{u}$ and $\frac{1}{v}$.\\
We mention it here because the same R-matrix (\ref{Rmatrix}) appears in spin chains and it is the building block of  the monodromy matrix $T(u)$ obeying the equation (\ref{rtt}). When the Hamiltonian of a spin chain can be obtained with a monodromy matrix, the system is integrable, in the sense that it is completely solvable. This is one of the reasons why the Yangian symmetry in $\cN = 4$ SYM has received a lot of attention in the last years, as we will discuss in the following.

\subsection{Representations of Yangian algebra}

Let us consider as an example $\gg = sl(m)$, which will be useful in discussing scattering amplitudes. 
An interesting representation $\pi$ satisfying all the properties listed above is the oscillator representation \cite{Drinfeld:1985rx,Drinfeld:1987sy}. The one-site level-zero generators are written in terms of the $W^A$ oscillator variables
\be
\pi(J^{A}{}_{B}) = W^{A} \frac{\partial}{\partial W^{B}} - \frac{\delta^A_B}{m}  h\,
\ee
where $A,B=1,\ldots,m$ and the operator $h$ is the generator of $U(1)$ in $gl(m)$:
\be
h =  \pi(J^{C}{}_{C}) = W^C\frac{\partial}{\partial W^C} \,,
\ee 
where summation over contracted indices is implicit.
The operator $h$ being central, we can decompose the space of functions of the $W^A$ into those of fixed degrees of homogeneity $h$. Thus we can think of $W$ as homogeneous coordinates on $\mathbb{C}\mathbb{P}^{m-1}$. Our representation acts on functions with fixed degrees of homogeneity on this space (which we will denote as $\mathcal{F}(\mathbb{CP}^{m-1})$).
The representation of the level-one generators can be obtained through the evaluation map \cite{Chari:1991xx}, which constructs the level-one operator $J^{(1)}$ in terms of the algebra generators $J$:
\begin{align}
\pi_\mu \bigl(J^{(1)}{}^{a}) &= \mu~ \pi(J^{a}) + \frac{1}{4} \sum_{bc} d^{a}_{~{bc}} \pi(J^{b} J^{c}) 
\end{align}
where $\mu$ is a free parameter and 
\be
d^{a}_{~{bc}} = \Tr\left[J^{a}\left(J_b J_c + J_c J_b\right)\right] \,.
\ee
We could compute this explicitly but there is a shortcut to the answer. In order to represent the level-one operator $J^{(1)}{}^{A}{}_{B}$ we need to write down an operator in the adjoint representation. Since the operator $h$ is central (and so can be assigned some fixed numerical value) our only choice is 
\be
\pi_\nu\left(J_\nu^{(1)}{}^{A}{}_{B}\right) = \nu \biggl( W^A \frac{\partial}{\partial W^B} - \tfrac{1}{m}\delta^A_B W^C \frac{\partial}{\partial W^C} \biggr) \,.
\ee
As we have previously mentioned,  we construct further representations acting on the tensor product $\mathcal{F}(\mathbb{CP}^{m-1}) \otimes \ldots \otimes \mathcal{F}(\mathbb{CP}^{m-1})$ by using the coproduct.
To construct now a two-parameter representation for two sites, we use (\ref{liecopr}) for the level zero
\begin{align}
\left(\pi_{\nu_1} \otimes \pi_{\nu_2}\right) (J^{A}{}_{B}) &=  \pi_{\nu_1} (J^{A}{}_{B}) \otimes 1 + 1 \otimes \pi_{\nu_2} (J^{A}{}_{B}) \\
&= W_1^{A} \frac{\partial}{\partial W_1^{B}} +  W_2^{A} \frac{\partial}{\partial W_2^{B}} -  \frac{\delta^A_B}{m} \sum_{i=1}^2   W_i^{C} \frac{\partial}{\partial W_i^{C}} 
\end{align}
and (\ref{coproduct}) for the level-one operators
\begin{align}
\left(\pi_{\nu_1} \otimes \pi_{\nu_2}\right) \bigl(J^{(1)}{}^{A}{}_{B}) &=  \pi_{\nu_1} \bigl(J^{(1)}{}^{A}{}_{B}) \otimes 1 + 1 \otimes \pi_{\nu_2} \bigl(J^{(1)}{}^{A}{}_{B}) +  f^{A~~F~D}_{~BE~C}   \pi_{\nu_1} (J^{C}{}_{D})  \otimes \pi_{\nu_2} (J^{E}{}_{F})\\
&=\nu_1 W_1^{A} \frac{\partial}{\partial W_1^{B}} + \nu_2  W_2^{A} \frac{\partial}{\partial W_2^{B}} -  \frac{\delta^A_B}{m} \sum_{i=1}^2  \nu_i  h_i^{C} \frac{\partial}{\partial W_i^{C}}  \nonumber \\
&+  \left( W_1^{A} \frac{\partial}{\partial W_1^{C}} W_2^{C} \frac{\partial}{\partial W_2^{B}} - (1,2)\right)\,,
\end{align}
where with $(i,j)$ we mean the exchange of the $i$ and $j$ indices. 
By  repeated application of the coproduct and the projection with $\pi_{\nu_i}$ on the $i$th site,  the representation for $n$ sites reads
\begin{align}
\pi_{\vec{\nu}}(J^{A}{}_{B}) &= \sum_{i=1}^n \biggl(W_i^A \frac{\partial}{\partial W_i^B} - \tfrac{1}{m}  \delta^A_B h_i\biggr)\,, \\
\pi_{\vec{\nu}}(J^{(1)}{}^{A}{}_{B}) &= \sum_{i<j} \biggl(W_i^A \frac{\partial}{\partial W_i^C}W_j^C \frac{\partial}{\partial W_j^B} - (i,j)\biggr) + \sum_{i=1}^n \nu_i \biggl(W_i^A \frac{\partial}{\partial W_i^B} -\tfrac{1}{m} \delta^A_B h_i\biggr)\,,
\label{birepr}
\end{align}
where $\vec{\nu} = (\nu_1,\ldots,\nu_n)$ is a list of free parameters. Equation (\ref{birepr}) is the bilocal formula (\ref{bilocal}). For the sake of simplicity, in the following we will use the symbols $J$ and $J^{(1)}_{\vec{\nu}}$ to denote this representation of the level-zero and level-one generators. Together the operators $J^{A}{}_{B}$ and $J_{\vec{\nu}}^{(1)}{}^{A}{}_{B}$ generate the Yangian $Y(sl(m))$.
The operators
\be
h_i = \ W_i^C \frac{\partial}{\partial W_i^C} 
\ee
are central and so we can decompose the space of functions of the $W_i$ into spaces of fixed homogeneity in each of the $W_i$ separately. 
In the following we will explicitly use these results in the context of scattering amplitudes.


\section{Scattering amplitudes, symmetries and Wilson loops}
\label{ampl}

In this section we describe some properties of (colour-ordered) scattering amplitudes in $\cN=4$ SYM. In particular we will focus on the symmetries of the theory and on the duality between Wilson loops and maximally-helicity violating  amplitudes.

\subsection{Superamplitudes and symmetries}

The on-shell supermultiplet of $\cN=4$ super Yang-Mills theory is conveniently described by a superfield $\Phi$, dependent on Grassmann parameters $\eta^A$ which transform in the fundamental representation of $su(4)$. The on-shell superfield can be decomposed as follows
\be
\Phi(p,\eta) = G^+(p) + \eta^A \Gamma_A(p) + \tfrac{1}{2!} \eta^A \eta^B S_{AB}(p) + \tfrac{1}{3!} \eta^A \eta^B \eta^C \e_{ABCD} \overline{\Gamma}^D(p) + \tfrac{1}{4!} \eta^A \eta^B \eta^C \eta^D \e_{ABCD} G^-(p).
\label{onshellmultiplet}
\ee
Here $G^+,\Gamma_A,S_{AB}=\tfrac{1}{2}\e_{ABCD}\overline{S}^{CD},\overline{\Gamma}^A,G^-$ are the positive helicity gluon, gluino, scalar, anti-gluino and negative helicity gluon states respectively. Each state $\phi \in \{G^+,\Gamma_A,S_{AB},\overline{\Gamma}^A,G^-\}$ carries a definite on-shell momentum. 
In the following we will use the spinor-helicity formalism, which is a very useful tool for describing scattering amplitudes of massless particles. The condition $p^2 = 0$ translates into writing the momentum as a bi-spinor
\be
\label{shform}
p^{\a \adt} = \lam^\a \tlam^\adt,
\ee
where $ \lam^\a$ and $\tlam^\adt$ transform in the (2, 0) and (0, 2) representations of the Lorentz group. 
The space defined by $( \lam^\a, \tlam^\adt, \eta^A)$ is called on-shell superspace. Each state $\phi$ 
has a definite weight $h$ (called helicity) under the rescaling
\be
\lam \longrightarrow \a \lam, \qquad \tlam \longrightarrow \a^{-1} \tlam, \qquad \phi(\lam,\tlam) \longrightarrow \a^{-2h} \phi(\lam,\tlam) \,.
\ee
The helicities of the states appearing in (\ref{onshellmultiplet}) are $\{+1,+\tfrac{1}{2},0,-\tfrac{1}{2},-1\}$ respectively. If, in addition, we assign $\eta$ to transform in the same way as $\tlam$,
\be
\eta^A \longrightarrow \a^{-1} \eta^A \,,
\ee
then the whole superfield $\Phi$ has helicity 1. In other words the helicity generator
\be
h = -\tfrac{1}{2} \lam^\a \frac{\del}{\del \lam^\a} + \tfrac{1}{2} \tlam^\adt \frac{\del}{\del \tlam^\adt} + \tfrac{1}{2} \eta^A \frac{\del}{\del \eta^A}
\ee
acts on $\Phi$ in the following way,
\be
h \Phi = \Phi.
\ee
When we consider scattering amplitudes of the on-shell superfields, then  the helicity condition (or `homogeneity condition') is satisfied for each particle, i.e.
\be
h_i \mathcal{A}_n(\Phi_1,\ldots,\Phi_n) = \mathcal{A}_n(\Phi_1,\ldots,\Phi_n), \qquad i=1,\ldots,n.
\label{Ahelicity}
\ee
The superamplitude $\mathcal{A}_n$ can be expanded in terms of the Grassmann parameters $\eta^A_i$ and written as follows
\be
\mathcal{A}_n =  \mathcal{A}_{n}^{\rm MHV} + \mathcal{A}_{n}^{\rm NMHV} + \ldots + \mathcal{A}_{n}^{\rm \overline{MHV}}=  \frac{\d^4(p) \d^8(q)}{\langle 12 \rangle \langle 23 \rangle \ldots \langle n1\rangle} \cP_n( \lam_i, \tlam_i, \eta_i) \,,
\label{amp}
\ee
where $\d^8\left(q\right)=\Pi_{A=1}^4 \Pi_{\alpha=1,2}\left(\sum_{i=1}^n \lam_{i \alpha} \eta_i^A\right)$ is a Grassmann delta function,  consequence of supersymmetry, and $\langle ij \rangle = \epsilon_{\alpha\beta} \lam_i^{\alpha} \lam_j^{\beta}$.
The function $\cP_n$, which will appear again later on, is a polynomial in the $\eta_i$'s and carries no helicity 
\be
h_i \mathcal{P}_n = 0, \qquad i=1,\ldots,n.
\label{Phelicity}
\ee
At tree level,  the function $\mathcal{P}_n$  is finite and its explicit form was given in \cite{Drummond:2008cr}. At loop level, it contains infrared divergences which need to be regularized. The regularisation procedure has the effect of breaking the symmetries of the theory, as we are going to discuss now.


Since $\cN =4$ SYM  is a superconformal field theory, one expects this to be reflected in the structure of the scattering amplitudes. This turns out to be true for tree-level amplitudes  but not at loop level, where the presence of infrared divergences spoils the symmetry. Therefore one should consider the invariance of the full S-matrix  under $psu(2,2|4)$ rather than of the amplitudes individually.
If we denote with $j_a$ any generator of the superconformal algebra $psu(2,2|4)$
\be
\label{jaconf}
j_a \in \{p^{\a\adt},q^{\a A}, \bar{q}^{\adt}_A,m_{\a\b}, \bar{m}_{\adt\bdt},r^A{}_B,d,s^\a_A,\bar{s}_{\adt}^A,k_{\a \adt} \}
\ee
then we can write at tree level
\be
j_a \mathcal{A}^{\mathrm{tree}}_n = 0. 
\label{scs}
\ee
The invariance was shown directly by applying the generators to the explicit form of the amplitudes in \cite{Witten:2003nn} for MHV amplitudes and \cite{Korchemsky:2009jv} for NMHV amplitudes.
In fact (\ref{scs}) is not completely exact, because of the so-called holomorphic anomaly  \cite{Bargheer:2009qu,Korchemsky:2009hm,Sever:2009aa}. This is given by special configurations of the external momenta, and specifically when two adjacent momenta become collinear. Even if at tree level this effect can be thought as negligible, since it is given just by some particular configuration, at loop level it must be taken into account due to the integration over the loop momentum running over the whole phase space. 
Moreover, at loop level there are effects due to the infrared regularisation procedure.
These effects can be taken into account by deforming the superconformal generators and in \cite{Bargheer:2009qu,Sever:2009aa,Beisert:2010gn} it was shown how to redefine them in such a way to restore symmetry at one loop.

But the ordinary superconformal symmetry  is not the full story. It has been revealed  \cite{Drummond:2008vq} that there exists also a hidden symmetry, which is not manifest in the on-shell superspace coordinates, the so-called dual superconformal symmetry with generators $J_a$. This second $psu(2,2|4)$ algebra is the T-dual version of the standard one, and this property corresponds very naturally to the  T-self-duality of the AdS sigma model \cite{Berkovits:2008ic,Beisert:2008iq,Beisert:2009cs}.  
The existence of such a symmetry was shown by defining a dual superspace $(x^{\a \adt}, \th^{\a A})$ through the relations
\be
x_i^{\a \adt} - x_{i+1}^{\a \adt} = \lam_i^\a \tlam_i^\adt, \qqquad \th_i^{\a A} - \th_{i+1}^{\a A} = \lam_i^\a \eta_i^A
\label{dualvars}
\ee
and re-expressing the amplitudes in terms of the dual coordinates
\be
\mathcal{A}_n = \frac{\d^4(x_1-x_{n+1}) \d^8(\th_1-\th_{n+1})}{\langle 12 \rangle \langle 23 \rangle \ldots \langle n1\rangle} \cP_n(x_i,\th_i).
\ee
In this space, dual superconformal symmetry acts canonically and it also acts on the on-shell superspace variables in order to be compatible with the constraints (\ref{dualvars}). In Appendix \ref{generators} we give the explicit expressions of the standard and dual superconformal generators. Some of the generators $J_a$ do not leave the amplitudes invariant. Indeed, under the dual superconformal $S^{\a A}$, special conformal $K^{\a \adt} $ and dilatation $D$ operators  the amplitudes are covariant  \cite{Drummond:2008vq, Brandhuber:2008pf}.
By simply redefining these generators \cite{Drummond:2009fd}, the covariance  can be rephrased as an invariance of $\mathcal{A}_n$ in such a way that at tree level
\be
J'_a \mathcal{A}^{\mathrm{tree}}_n = 0 \,,
\ee
where $J'_a$'s are any generator of the dual copy of $psu(2,2|4)$
\be
J'_a \in \{P_{\a\adt},Q_{\a A}, \bar{Q}_{\adt}^A,M_{\a\b}, \overline{M}_{\adt\bdt},R^A{}_B,D',S_\a^{\prime A},\overline{S}^{\adt}_A,K'^{\a \adt} \}.
\ee
The redefined generators still satisfy the commutation relations of the superconformal algebra, but now with vanishing central charge, $C' = 0$.
In order to have both symmetries acting on the same space, it is useful to restrict the dual superconformal generators to act only on the on-shell superspace variables $(\lam_i,\tlam_i,\eta_i)$. Then one finds that the generators $P_{\a \adt},Q_{\a A}$ become trivial, while the generators $\{\bar{Q},M,\bar{M},R,D',\bar{S}\}$ coincide (up to signs) with generators of the standard superconformal symmetry. The non-trivial generators which are not part of the $j_a$ are $K'$ and $S'$.

In \cite{Drummond:2009fd} it was shown that the generators $j_a$ (\ref{jaconf}) and $S'$ (or $K'$) together generate the Yangian of the superconformal algebra, $Y(psu(2,2|4))$. As we have already commented, the presence of a Yangian structure suggests once more the integrability of $\NN=4$ SYM. This had already been pointed out from the evaluation of the spectrum of anomalous dimensions \cite{AdS/INT1,AdS/INT1b,Beisert:2005fw} and, through the AdS/CFT correspondence, in the classically integrability  of IIB superstring on  AdS${}_5 \times$S${}^5$ \cite{Bena:2003wd}.\\
In particular, the generators $j_a$ form the level-zero $psu(2,2|4)$ subalgebra as in (\ref{levelzero}),
while the level one is generated by considering $S'$ or $K'$ in (\ref{levelone}).  
Let us stress that the choice of $S'$ or $K'$ does not change the Yangian, since $[K', \bar{Q}] = S'$.
The level-zero generators are represented by the sum over single particle generators (\ref{sumzero}),
\be
j_a = \sum_{k=1}^n j_{ka} \,,
\ee
while the level-one generators are obtained via the bilocal formula (\ref{bilocal})
\be
j_a\!{}^{(1)} = f_{a}{}^{cb} \sum_{k<k'} j_{kb} j_{k'c} \,,
\label{bilocal1}
\ee
where for the time being the parameters $c_k$ can be set to zero.
In  \cite{Dolan:2004ps}  it was demonstrated that under certain conditions, which turn out to be satisfied by the oscillator representation in exam,  the formula (\ref{bilocal1}) is sufficient to define the Yangian, so that the Serre relations are automatically satisfied. The bilocal formula also respects the cyclicity of the amplitudes for the case   of $psl(n|n)$.
Thus finally the full symmetry of the tree-level amplitudes can be rephrased as
\be
y \mathcal{A}^{\mathrm{tree}}_n = 0,
\ee
for any $y \in Y(psu(2,2|4))$.
As we have already mentioned, at tree level there are correction terms of the superconformal symmetry given by the holomorphic anomaly. With these deformations, the algebra  (\ref{levelzero}) does close only modulo gauge transformations \cite{Bargheer:2009qu}. By using the bilocal formula, the deformations are directly inserted in a new definition of level-one generators.   At loop level, one can again define deformed level-zero and -one generators  \cite{Sever:2009aa,Beisert:2010gn}. In both tree- and loop-level cases,  it has not been checked that the deformed generators close into a Yangian algebra.

In \cite{Drummond:2010qh} it was demonstrated that  there exists an alternative representation of the symmetry, which can be viewed as the Yangian version of the T-self-duality property of the full AdS${}_5\times$S${}^5$ background of the string sigma-model \cite{Berkovits:2008ic,Beisert:2008iq,Beisert:2009cs}.
In this case, it is the unmodified dual superconformal symmetry $J_a$ which plays the role of level-zero subalgebra. Now it is  the function $\cP_n$, so the amplitude with the MHV part factored out, to be left invariant at tree level
\be
J_a \mathcal{P}^{\mathrm{tree}}_n = 0 \,.
\ee
The level one is then provided by some of the (modified) standard superconformal generators. Similarly to the previous case, the generators of translations $p^{\a\adt}$ and supertranslations $q^{\a A}$ are trivialised while 
$k_{\a \adt}$ and $s^\a_A$ are redefined in such a way to annihilate $\cP_n^{\mathrm{tree}}$. Then $J_a$ and $k'$ (or $s'$) generate the Yangian of the dual superconformal algebra.
\begin{figure}[h]
\psfrag{j}[cc][cc]{$j$} \psfrag{J}[cc][cc]{$J$}
\psfrag{j1}[cc][cc]{$j^{(1)}$} \psfrag{J1}[cc][cc]{$J^{(1)}$}
\psfrag{pq}[cc][cc]{\!\!\!\!\!\!\!\!\!\!\!$p,q$}\psfrag{PQ}[cc][cc]{\,\,\,\,$P,Q$}
\psfrag{KS}[cc][cc]{\!\!\!\!\!\!\!\!\!\!\!$K,S$}\psfrag{ks}[cc][cc]{\,\,\,\,$k,s$}
\psfrag{rR}[cc][cc]{}
\psfrag{Tduality}[cc][cc]{T-duality}
\centerline{{\epsfysize9cm \epsfbox{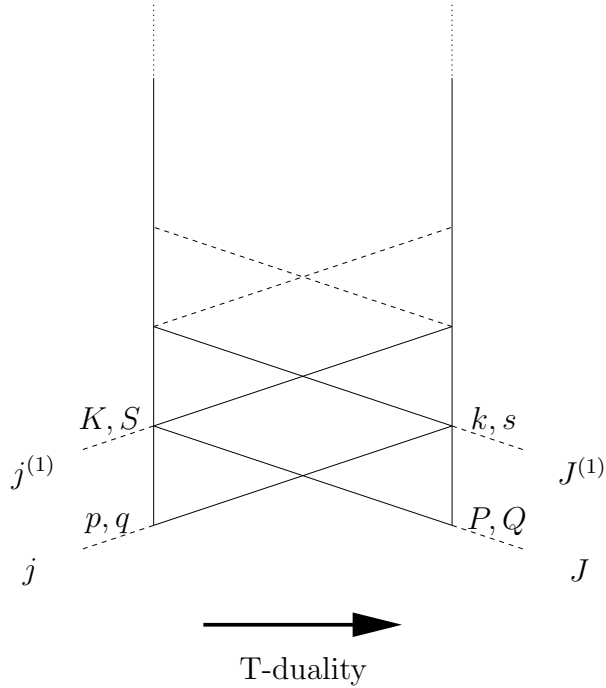}}} \caption[]{\small The tower of symmetries acting on scattering amplitudes in $\cN=4$ super Yang-Mills theory. The original superconformal charges are denoted by $j$ and the dual ones by $J$. Each can be thought of as the level-zero part of the Yangian $Y(psu(2,2|4))$. The dual superconformal charges $K$ and $S$ form part of the level-one $j^{(1)}$ while the original superconformal charges $k$ and $s$ form part of the level one charges $J^{(1)}$. In each representation the `negative' level ($P$ and $Q$ or $p$ and $q$) is trivialised. T-duality maps $j$ to $J$ and $j^{(1)}$ to $J^{(1)}$.}
\label{Fig:aux}
\end{figure}
Therefore there are two equivalent ways of looking at the full symmetry algebra of the scattering amplitudes. We summarize them in Table \ref{tab:twtwistor}, where we introduce the twistors $\mathcal{Z}^{\AA}$ and momentum twistors $\mathcal{W}^{\AA} $. The formulation in these spaces has a particularly simple form  \cite{Drummond:2009fd,Drummond:2010qh}, and in particular the level-zero generators are linearised in terms of these variables in both spaces. Moreover, the identical formulations show explicitly the T-self-duality property. 
The twistor coordinates are defined as
$\mathcal{Z}^{\AA} = (\tilde{\mu}^{\alpha}, \tilde\lambda^{\dot\alpha} , \eta^A)$, where $\tilde\mu$ is the Fourier conjugate  of $\lambda$. The momentum twistors are instead $\mathcal{W}^{\AA} = (\lambda^{\alpha}, \mu^{\dot\alpha} , \chi^A)$, where $\mu^{\dot\alpha}$ and $\chi^A$ are defined through the incidence relations
\be
\mu_i^{\adt} = x_i^{\a \adt} \lam_{i \a}, \qqquad \chi_i^A = \th_i^{\a A}\lam_{i \a} \,.
\label{incidence}
\ee
This is the standard twistor space associated to Wilson loops and was recently introduced in \cite{Hodges:2009hk}.
One of the most useful properties of the momentum twistors is that the lightlikeness constraint and momentum conservation  are already implemented via the incidence relations (\ref{incidence}). 
\begin{table}[t]
\begin{tabular}{ c | c }
Twistor space & Momentum twistor space \\ \hline
 ~ & ~ \\
 $\displaystyle  j^{\AA}{}_{\BB} = \sum_i \cZ_i^{\AA} \frac{\del}{\del \cZ_i^{\BB}}  \label{twistorsconf} $ & $\displaystyle J^{\AA}{}_{\BB} = \sum_i \cW_i^{\AA} \frac{\del}{\del \cW_i^{\BB}}$  \\  
 ~ & ~ \\
$ \displaystyle j^{(1)}{}^{\AA}{}_{\BB} = \sum_{i<j} (-1)^{\CC}\Bigl[\cZ_i^{\AA} \frac{\del}{\del \cZ_i^{\CC}} \cZ_j^{\CC} \frac{\del}{\del \cZ_j^{\BB}} - (i,j) \Bigr]$ & $ \displaystyle J^{(1)}{}^{\AA}{}_{\BB} = \sum_{i<j} (-1)^{\CC}\Bigl[\cW_i^{\AA} \frac{\del}{\del \cW_i^{\CC}} \cW_j^{\CC} \frac{\del}{\del \cW_j^{\BB}} - (i,j) \Bigr]$  \\ 
~ & ~ \\
$\displaystyle j \mathcal{A}_n = j^{(1)} \mathcal{A}_n = 0$ & $\displaystyle J \mathcal{P}_n = J^{(1)} \mathcal{P}_n = 0$ \\ 
 ~ & ~ \\
\end{tabular}
\caption{Yangian generators in twistor and momentum twistor coordinates.}
\label{tab:twtwistor}
\end{table}
\\The T-dual relationship between twistor and momentum twistor space is exactly the property which appears in the Grassmannian formulas of \cite{ArkaniHamed:2009dn} and \cite{Mason:2009qx}, as we will explain in section \ref{Yinvariants}.

Before moving to the Wilson loop/amplitude duality, we want to point out some references about recent results related to the topics discussed above. 
In particular, we want to mention that there is another alternative regularisation which preserves  the dual conformal symmetry at loop level  \cite{Alday:2009zm,Henn:2010bk,Henn:2010ir} by going to the Coulomb branch of the theory. 
See \cite{Bern:2010qa,CaronHuot:2010rj,Dennen:2010dh} for higher dimensional related theories. 
Another very interesting result is  the enhancement of the  $psu(2,2|4)$ algebra 
through the hypercharge generator $\mathfrak{B}$, which measures the helicity of scattering amplitudes. While $\mathfrak{B}$ is broken, the level-one helicity generator associated to it is a symmetry of the amplitudes in $\cN =4$ SYM. See \cite{Beisert:2011pn} for more details.


\subsection{MHV amplitude/Wilson loop duality}
\label{MHV_WL}

In the following we will describe the duality between MHV amplitudes and Wilson loops, which will be useful for the last section.  For some recent papers  about the extension of the duality to non-MHV amplitudes at weak coupling see \cite{Belitsky:2011zm,Mason:2010yk,CaronHuot:2010ek}. 

The MHV  amplitude is one term of the superamplitude $\mathcal{A}_n$ when expanded in terms of the Grassmann variables $\eta_i$'s (\ref{amp}) and can be written as follows:
\be
 \mathcal{A}_{n}^{\rm MHV} =  \mathcal{A}_{n;tree}^{\rm MHV} M_n \,.
 \label{MHV}
\ee
Like $\cP_n$, the function $M_n$ is the loop-correction factor and does not carry any helicity, that is completely encoded in the prefactor. The exponentiation of infrared divergences leads to write 
\be
\log M_n =  [\text{IR divergences}] + F_n^{\rm MHV}(p_1,\ldots,p_n; a) + O(\epsilon)\,
\label{Mn}
\ee
where $a = \frac{g^2 N}{8 \pi^2}$ is the 't Hooft coupling.
The first term on the RHS, encoding the infrared divergences, is well understood in $\cN =4$ SYM. In particular, the leading IR singularity is a double pole, proportional to the cusp anomalous dimension. This quantity is known to be related to ultraviolet divergences of Wilson loops with cusps in QCD \cite{Ivanov:1985bk,Ivanov:1985np,Korchemsky:1985xj}.
The finite part is instead under strong analysis in the last years, and many developments in its computation have been achieved recently. Indeed in  \cite{Bern:2005iz} Bern, Dixon and Smirnov (BDS) proposed an ansatz for $F_n^{\rm MHV}$ based on calculations of the four-point amplitude and verified to hold for five-point scatterings.  In the meanwhile, using the AdS/CFT correspondence, in \cite{Alday:2007hr} it was showed that  in the strong coupling regime there is a relation between planar $n$-gluon scattering amplitudes (the dependence on the helicity configuration of the amplitude appears as a subleading effect at strong coupling) and Wilson loops defined on closed light-like contours $C_n$:
\begin{equation}
W_n =  \langle 0| \mathrm{\cP~exp} \left(ig \oint_{C_n}  A\right) |0 \rangle .
\end{equation}
This means that not only the IR divergences of the scattering amplitude match the UV divergences of the light-like Wilson loop given by the cusps. It also  implies that the finite parts have to coincide.
Then it was  conjectured and verified that such a duality exists also in the perturbative regime for MHV amplitudes \cite{Drummond:2007aua,Brandhuber:2007yx,Drummond:2007cf,Drummond:2007au,Drummond:2007bm,Drummond:2008aq}.
It formally arises from the identification of particle momenta with the dual coordinates (\ref{dualvars})
\be
x_i - x_{i+1} = p_i, \qqquad  (x_i - x_{i+1})^2 = 0 \,,
\label{dualvarsbos}
\ee
and a formula similar to (\ref{Mn}) can be written for the Wilson loops in dual space:
\be
\log W_n = [\text{UV divergences}] + F_n^{\rm WL}(x_1,\ldots,x_n; a) + O(\epsilon)\,.
\label{Wn}
\ee
Therefore the Wilson loop $W_n$ can be seen as the MHV scattering amplitude with the tree-level part factored out, upon going from on-shell to dual coordinates.
The duality then is really a statement on the finite parts (up to an additive constant)
\begin{equation}
F_n^{\mathrm{MHV}} = F_n^{\mathrm{WL}} + \mathrm{const} \,,
\end{equation}
upon the identification (\ref{dualvarsbos}).
The dual conformal symmetry for the scattering amplitudes, whose supersymmetric formulation we have discussed in the previous section, is the ordinary conformal symmetry of Wilson loops. The dual space is the configuration space for the Wilson loops and for this reason, the breaking of the symmetry is under control through anomalous Ward identities. In particular it has been found  \cite{Drummond:2007cf,Drummond:2007au} that
\be
K^\mu F_n^{\rm WL}(x_1,\ldots,x_n) = \frac{1}{2} \Gamma_{\rm cusp}(a) \sum_{i=1}^n x_{i,i+1}^\mu \log \frac{x_{i,i+2}^2}{x_{i-1,i+1}^2}\,.
\ee
 Using these Ward identities, it was shown that the BDS ansatz can differ from the Wilson loop result by a non-trivial function of $(3n -15)$ conformal invariants $u_{ijkl}$ 
\begin{equation}
u_{ijkl} = \frac{x_{ij}^2 x_{kl}^2}{x_{ik}^2 x_{jl}^2} \,.
\label{cross}
\end{equation}
Since the first case where the conformal cross-ratios  (\ref{cross}) can appear is at six points,  the BDS ansatz, centred upon four- and five-gluons amplitudes,  did not detect  this term.
The discrepancy, called remainder function $\mathcal{R}$, has been verified to be non-zero at two loops and beyond, and for six or more points \cite{Drummond:2007bm}-\cite{CaronHuot:2011ky}.
Thus we can write the finite part as the sum of two contributions
\be
\label{Fanom}
F_n^{\mathrm{WL}} =  F_n^{\rm anom}(x_1,\ldots,x_n; a) + I_n(u_1,\ldots,u_m; a) \,.
\ee
If we choose $F_n^{\rm anom}$ to coincide with the BDS ansatz for the MHV amplitude, then $I_n$ coincide with  the standard definition of 
$\mathcal{R}$.
But there are alternative definitions of the anomalous part which modify it by adding some function of invariants and subtracting the same function from the remainder function.
We will delve into this in the last section.


\section{Yangian invariants}
\label{Yinvariants}

We want now to address the issue of Yangian invariants. We will begin by reviewing  recent results on the connection between Grassmanians, Yangian invariants and scattering amplitudes. Then  we will focus on a particular function constructed through Wilson loops, which turns out to be Yangian invariant at one loop.

\subsection{Superamplitudes and Grassmannians}

In \cite{ArkaniHamed:2009dn}  a remarkable formula was proposed which computes leading singularities of N${}^{k-2}$MHV scattering amplitudes in the $\NN=4$ super Yang-Mills theory. It was formulated in twistor space and takes the form of an integral over the Grassmannian $G(k,n)$, the space of complex $k$-planes in $\mathbb{C}^n$, of certain superconformally invariant delta functions
\be
\mathcal{L}_{\rm ACCK}(\cZ) = \int \frac{ D^{k(n-k)}c}{\MM_1 \ldots \MM_n} \prod_{a=1}^k \delta^{4|4}\Bigl(\sum_{i=1}^n c_{ai} \cZ_i\Bigr) .
\label{ACCK}
\ee
Here the $c_{ai}$ are complex parameters which are integrated choosing a specific contour.
The denominator is the cyclic product of  consecutive $(k \times k)$ minors $\MM_p$  made from the columns $
p,\ldots,p+k-1$ of the $(k \times n)$ matrix of the $c_{ai}$
\beq
\MM_{p} \equiv (p~p+1~p+2\ldots p+k-1)  .
\eeq 	
The integrand is a specific $k(n-k)$-form to be integrated over cycles, with the integral being treated as a multi-dimensional contour integral.
The result obtained, and therefore the leading singularities, depends on the choice of contour and is non-vanishing for closed contours because the form has poles located on certain hyperplanes in the Grassmannian.
The formula (\ref{ACCK}) has a T-dual version \cite{Mason:2009qx}, expressed in terms of momentum twistors. The momentum twistor Grassmannian formula takes the same form as the original
\be
\mathcal{L}_{\rm MS}(\cW) = \int \frac{D^{k(n-k)}t}{\MM_1 \ldots \MM_n} \prod_{a=1}^k  \delta^{4|4}\Bigl(\sum_{i=1}^n t_{ai} \cW_i\Bigr) ,
\label{MS}
\ee
but now it is the dual superconformal symmetry that is manifest. The integration variables $t_{ai}$ are again a $(k \times n)$ matrix of complex parameters. The formula (\ref{MS}) produces the same objects as (\ref{ACCK}) but now with the MHV tree-level amplitude factored out. Thus they contribute to N${}^k$MHV amplitudes.
The equivalence of the two formulations (\ref{ACCK}) and (\ref{MS}) was shown in \cite{ArkaniHamed:2009vw}, through a change of variables. Therefore, since each of the formulas has a different superconformal symmetry manifest, they both possess an invariance under the Yangian $Y(psl(4|4))$. The Yangian symmetry of these formulas was explicitly demonstrated in \cite{Drummond:2010qh} by directly applying the Yangian level-one generators to the Grassmannian integral itself. 
In particular  it was shown that $J^{(1)}{}^{\AA}{}_{\BB}$ on $\mathcal{L}_{\rm MS}(\cW)$  yields a total derivative,
which guarantees that $\mathcal{L}$ is invariant when the contour is closed. 
It was later shown \cite{Drummond:2010uq,Korchemsky:2010ut}  that the form of the Grassmannian integral is uniquely fixed by requiring Yangian invariance. 
Specifically, in \cite{Drummond:2010uq}, by using the methods developed in \cite{Drummond:2010qh}, 
it was demonstrated that one cannot modify the integrand by a non-constant multiplicative function without breaking Yangian invariance (in the case of zero-homogeneity condition).

At loop level it has recently been realised that the above statements all hold at the level of the (unregularized) all-loop planar integrand  \cite{ArkaniHamed:2010kv}. The integrand at a given loop order has been constructed from its singularities via a generalisation of the BCFW recursion relations, whose each term is individually invariant under the full Yangian symmetry up to a total derivative. But at the level of the actual amplitudes the situation with the full symmetry is less clear. The Yangian invariance is broken by the action of integrating over specific contours, which leads to infrared divergences.\\
There has been a very recent result \cite{Drummond:2010zv} for the Yangian invariance at one loop for Wilson loops (or equivalently MHV amplitudes), as we are going to describe now.

\subsection{Wilson loops and the ratio function}

\begin{figure}
 \centerline{{\epsfysize4.5cm
\epsfbox{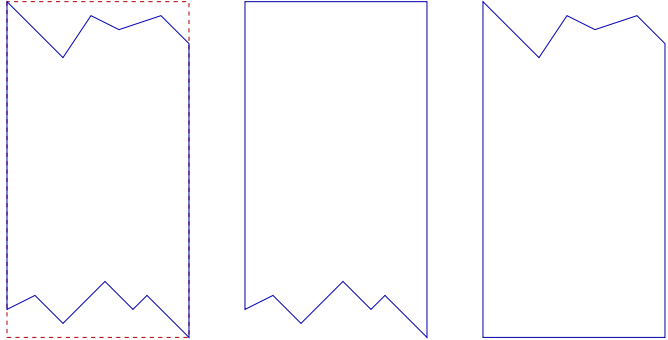}}}  \caption[]{\small The four different Wilson loops entering the definition of the ratio (\ref{ratio}). The reference square is shown by the dashed line. The bottom and top Wilson loops are obtained by replacing a sequence of edges by the corresponding part of the square.}
  \label{Wloops}
\end{figure}

In section (\ref{MHV_WL}) we have reviewed the duality between MHV scattering amplitudes and Wilson loops. As we have pointed out, much effort has been put on evaluating the finite part $F_n^{WL}$, as the divergent piece of Wilson loops is already well understood.
However the definition of $F_n^{\rm anom}$ in  (\ref{Fanom}) is ambiguous because it can be modified by any function of the available conformal invariants.  An example is the definition of the `BDS-like' piece of \cite{Alday:2009yn} where the anomalous part depends only on the shortest distances $x_{i,i+2}^2$.
A particularly interesting definition for the decomposition was made in \cite{Alday:2010ku}. In this case one picks two of the light-like edges and forms a light-like square by picking two more light-like lines intersecting them both.  Then one can consider four different Wilson loops. The original Wilson loop $W_n$, the Wilson loop on the square $W_{\rm sq}$ and the Wilson loops formed by replacing the top or bottom set of intermediate edges by the corresponding part of the square, $W_{\rm top}$ and $W_{\rm bottom}$ respectively. 
This is best illustrated by Fig. \ref{Wloops}. 
A conformally invariant quantity, function  of the cross-ratios $u_1,\ldots,u_{k}$, can be defined by the following ratio of the Wilson loops
\be
r_n = \log \biggl(\frac{W_n W_{\rm sq}}{W_{\rm top} W_{\rm bottom}}\biggr).
\label{ratio}
\ee
Unlike the usual definition of the remainder function $\mathcal{R}$, $r_n$ is non-zero already at one loop. It is also not cyclic invariant since its definition requires a choice of two special lines from which to form the square. This choice essentially corresponds to the choice of OPE `channel' in which one expands the Wilson loop over exchanged intermediate excited flux tube states \cite{Alday:2010ku}.
The quantity $r_n$ is particularly simple at one loop. It corresponds to the connected part of the correlation between the two Wilson loops. 
A further simplification is obtained when considering restricted two-dimensional kinematics as in \cite{Alday:2009yn}. In this case one needs an even number of sides to the Wilson loops, alternating in orientation between the $x^+$ direction and the $x^-$ direction as one travels round the loop. The number of independent cross-ratios is reduced in the two-dimensional kinematics. In fact there are $(n-6)$ independent ratios left from the original $(3n-15)$. Since $n$ is always even, the first non-trivial ratio is therefore at eight points.
\\As we are delving with Wilson loops, thus with the function $M_n$ defined in (\ref{MHV}), we consider the $sl(4)$ bosonic subspace of momentum twistors $W_i^{A} = (\lambda^{\alpha}, \mu^{\dot\alpha})$ and functions with no helicity
\be
h_i = W_i^C\frac{\partial}{\partial W_i^C} = 0 \,.
\ee
Actually, when restricting to the special two-dimensional kinematics,  the twistor variables are also restricted and preserve two commuting copies of $sl(2)$ inside $sl(4)$.
Specifically we can decompose the $W_i^A$ into upper and lower components each transforming under its own  $sl(2)$. The alternating orientations of the lines correspond to an alternating between twistors transforming under the two copies of $sl(2)$. We take the odd-numbered twistors to transform under the first copy and the even-numbered ones to transform under the second copy,
\be
W_{2i+1} = 
\left(
\begin{matrix} w_{2i+1}\\0
\end{matrix}
\right),
\qquad
W_{2i} = 
\left(
\begin{matrix} 0 \\ \bar{w}_{2i}
\end{matrix}
\right)\,.
\ee
In this two-dimensional case the cross-ratios are simply 
\be
u = \frac{(15) (37)}{(17)(35)} , \qquad v = \frac{(26)(48)}{(28)(46)} \,,
\ee
with 
$
(i j) =  \epsilon_{A_1  A_2} W_{i}^{A_1} W_{j}^{A_2}\,
$.
The ratio $r_n$ is simple to compute since the relevant Wilson loops are known at one loop \cite{Brandhuber:2007yx} to coincide with the one-loop MHV amplitudes \cite{Bern:1994zx} through the duality. 
In the special two-dimensional kinematics all quantities can be expressed in terms of logarithms \cite{Alday:2009yn,Heslop:2010kq}.
Remarkably the function $r_8$ at one loop is none other than \cite{Gaiotto:2010fk}
\be
\label{r82d}
r_8 = g^2 \log u \log v + {\mathrm{const}} \,.
\ee
Also the Yangian symmetry is restricted to the $Y(sl(2)) \oplus Y(sl(2))$ subalgebra.
Following section \ref{Yalgebra}, in particular the formulas (\ref{sumzero}) and (\ref{bilocal}), we can write the generators of the two copies of $Y(sl(2))$ as (we recall that $h_i=0$)
\begin{align}
J^{a}{}_{b} &= \sum_{i \text{ odd}}  w_i^a \frac{\partial}{\partial w_i^b} \,, \qquad J^{(1)}{}^{a}{}_{b} = \sum_{i<j \text{ odd}} \biggl(w_i^a \frac{\partial}{\partial w_i^c} w_j^c \frac{\partial}{\partial w_j^b} - (i,j)\biggr) + \sum_{i \text{ odd}} \nu_i w_i^a \frac{\partial}{\partial w_i^b}  \\
\overline{J}^{\bar{a}}{}_{\bar{b}} &=  \sum_{i \text{ even}} \bar{w}_i^{\bar{a}} \frac{\partial}{\partial \bar{w}_i^{\bar{b}}} \,,
  \qquad \overline{J}^{(1)}{}^{\bar{a}}{}_{\bar{b}} =  \sum_{i<j \text{ even}} \biggl(\bar{w}_i^{\bar{a}} \frac{\partial}{\partial \bar{w}_i^{\bar{c}}} \bar{w}_j^{\bar{c}} \frac{\partial}{\partial \bar{w}_j^{\bar{b}}} - (i,j)\biggr) + \sum_{i \text{ even}} \bar{\nu}_i \bar{w}_i^{\bar{a}} \frac{\partial}{\partial \bar{w}_i^{\bar{b}}} \,.
  \label{sl2s}
\end{align}
Here $a,b$ and $\bar{a},\bar{b}$ run from 1 to 2 and the $\nu_i, \bar{\nu}_i$'s are the free parameters we had indicated by $c_k$ in (\ref{bilocal}).
Now let us consider functions of the $w_i$ which are invariant under the action of the Yangian generators. Let us start with the level-zero conformal algebra. As we are just interested in $sl(2)$-invariant functions of the $w_i$, we can have any function of the invariant quantities,
\be
(i_1 i_2) =  \epsilon_{a_1  a_2} w_{i_1}^{a_1} w_{i_2}^{a_2}\,.
\label{slninvariants}
\ee 
By requiring homogeneous functions with degree zero in all of the $w_i$, we must consider functions of homogeneous ratios of the invariants in equation (\ref{slninvariants}). The first possibility to form a homogeneous ratio is at four sites (1,2,3,4) where we can write
\be
u=\frac{(13)(24)}{(14)(23)}\,,
\ee
which is the only independent invariant. The only other possibility is related to $u$ using the cyclic identity $(ij)w_k^a + (jk)w_i^a + (ki)w_j^a =0$,
\be
\frac{(12)(34)}{(41)(23)} = 1-u\,.
\ee
Thus the $sl(2)$ invariant functions on four copies of $(\mathbb{CP}^1)$ are functions of $u$. Requiring that they are also Yangian invariant functions means that we have to solve the equations
\be
J_{\vec{\nu}}^{(1)}{}^{a}{}_{b} f(u) = 0\,.
\ee
Obviously a constant function is always a solution of the equations.
There exists also a non-trivial one-parameter solution given by  hypergeometric functions,
\be
f_{\mu}(u) = \frac{(1-u)^{1+\mu}}{1+\mu}\,\, {}_2F_1(1,1+\mu,2+\mu;1-u)\,,\qquad  \mu=\tfrac{1}{2}(\nu_2-\nu_1)\,,
\ee
with the constraint  $\nu_1-\nu_3=\nu_2-\nu_4=2$.  These functions represent the only homogeneous functions at four sites which are also Yangian invariants.  If we consider the special case $\mu=0$, whose related values of the weights $\nu_i$'s correspond to demanding cyclicity  of the four-site level-one generators, 
we get:
\be
f_0(u) = (1-u) \,\,{}_2F_1(1,1,2;1-u) = \log u\,,
\ee
which is exactly one of the terms in (\ref{r82d}) after renumbering of the sides.
Thus finally we can write\footnote{For the details of the computation see \cite{Drummond:2010zv}.}
\be
J^{a}{}_{b} \log u= 0\,, \qquad J^{(1)}_{\vec{\nu}}{}^{a}{}_{b} \log u= 0\,, \qquad \vec{\nu}=(1,1,-1,-1)\,.
\ee
Therefore $r_8$ is Yangian invariant under two copies of the Yangian $Y(sl(2))$ for the choices $\vec{\nu}=(1,1,-1,-1)$ and $\vec{\bar{\nu}} = (1,1,-1,-1)$ where the entries range over the odd and even values of the light-like segments respectively.
Moreover, as pointed out in \cite{Gaiotto:2010fk}, at one loop there is a very simple relation between $r_n$ and $r_{n-2}$ for a given choice of reference square. This amounts to the fact that at one loop the Wilson loops are additive in nature. We can use this reduction argument to write
\beq
r_{n} = f(r_{n-2})=g(r_{n-4})=...=h(r_8)  \,.
\eeq
Thus we conclude the $r_{n}$ is always invariant under the two commuting Yangians with a natural representation corresponding to the choice of OPE channel (i.e. choice of reference square).

This analysis leads us to one of the most urgent questions in this respect. It is indeed the first indication that the Yangian symmetry seen at the level of the tree amplitudes (or the integrand for loop amplitudes) exhibits itself in a simple and natural way on the functions at one loop. 
Of course in the case presented, there are only two copies of the bosonic Yangian $Y(sl(2))$, not the full $Y(psl(4|4))$. The fate of the symmetry beyond one loop and the special momentum configuration is still unknown. However  the above results are a strong indication that it should be still present in the general case.


\subsection*{Acknowledgements}

I would like to thank James Drummond and Eric Ragoucy for collaboration on these topics and useful comments on the review.

\appendix

\setcounter{section}{0} \setcounter{equation}{0}
\renewcommand{\theequation}{\Alph{section}.\arabic{equation}}

\section{Ordinary and dual superconformal generators}
\label{generators}

We list here the explicit expressions of the generators of the conventional and dual superconformal symmetries.
We will use the following shorthand notation:
\begin{align}\label{shortderiv}
\partial_{i \alpha \dot{\alpha}} = \frac{\partial}{\partial
x_i^{\alpha \dot{\alpha}}}, \qquad \partial_{i \alpha A} = \frac{\partial}{\partial \theta_i^{\alpha
A}}, \qquad \partial_{i \alpha} = \frac{\partial}{\partial \lambda_i^{\alpha}}\,, \qquad
\partial_{i \dot{\alpha}} = \frac{\partial}{\partial
    \tilde{\lambda}_i^{\dot{\alpha}}}\,, \qquad
\partial_{i A} = \frac{\partial}{\partial \eta_i^A}\,.
\end{align}
The generators of the conventional superconformal symmetry are
\begin{align}
& p^{\dot{\alpha}\alpha }  =  \sum_i \tilde{\lambda}_i^{\dot{\alpha}}\lambda_i^{\alpha} \,, & &
k_{\alpha \dot{\alpha}} = \sum_i \partial_{i \alpha} \partial_{i \dot{\alpha}} \,,\notag\\
&\overline{m}_{\dot{\alpha} \dot{\beta}} = \sum_i \tilde{\lambda}_{i (\dot{\alpha}} \partial_{i
\dot{\beta} )}, & & m_{\alpha \beta} = \sum_i \lambda_{i (\alpha} \partial_{i \beta )}
\,,\notag\\
& d =  \sum_i [\tfrac{1}{2}\lambda_i^{\alpha} \partial_{i \alpha} +\tfrac{1}{2}
\tilde{\lambda}_i^{\dot{\alpha}} \partial_{i
    \dot{\alpha}} +1], & & r^{A}{}_{B} = \sum_i [-\eta_i^A \partial_{i B} + \tfrac{1}{4}\delta^A_B \eta_i^C \partial_{i C}]\,,\notag\\
&q^{\alpha A} =  \sum_i \lambda_i^{\alpha} \eta_i^A \,, &&   \bar{q}^{\dot\alpha}_A
= \sum_i \tilde\lambda_i^\adt \partial_{i A} \,, \notag\\
& s_{\alpha A} =  \sum_i \partial_{i \alpha} \partial_{i A}, & &
\bar{s}_{\dot\alpha}^A = \sum_i \eta_i^A \partial_{i \dot\alpha}\,,\notag\\
&c = \sum_i [1 + \tfrac{1}{2} \lambda_i^{\a} \partial_{i \a} - \tfrac{1}{2} \tilde\lambda^{\adt}_i \partial_{i \adt} - \tfrac{1}{2} \eta^A_i \partial_{iA} ]\,,
\end{align}
where with $(\alpha \beta)$ we mean the indices symmetrized.
The generators of dual superconformal transformations can be constructed by starting with the standard
chiral representation and extending the generators so that they commute with the constraints,
\be
(x_i-x_{i+1})_{\a \dot\alpha}  - \lambda_{i\, \a}\, \tilde{\lambda}_{i\, \dot\alpha} = 0\,, \qquad (\theta_i - \theta_{i+1})_\alpha^A - \lambda_{i \alpha} \eta_i^A = 0\,.
\ee
By construction they preserve the surface defined by these constraints, which is where the amplitude
has support. The generators are
\begin{align}
P_{\alpha \dot{\alpha}}&= \sum_i \partial_{i \alpha \dot{\alpha}}\,, \qquad Q_{\alpha A} = \sum_i \partial_{i \alpha A}\,, \qquad
\overline{Q}_{\dot{\alpha}}^A = \sum_i [\theta_i^{\alpha A}
  \partial_{i \alpha \dot{\alpha}} + \eta_i^A \partial_{i \dot{\alpha}}], \notag\\
M_{\alpha \beta} &= \sum_i[x_{i ( \alpha}{}^{\dot{\alpha}}
  \partial_{i \beta ) \dot{\alpha}} + \theta_{i (\alpha}^A \partial_{i
  \beta) A} + \lambda_{i (\alpha} \partial_{i \beta)}]\,, \qquad
\overline{M}_{\dot{\alpha} \dot{\beta}} = \sum_i [x_{i
    (\dot{\alpha}}{}^{\alpha} \partial_{i \dot{\beta} ) \alpha} +
  \tilde{\lambda}_{i(\dot{\alpha}} \partial_{i \dot{\beta})}]\,,\notag\\
R^{A}{}_{B} &= \sum_i [\theta_i^{\alpha A} \partial_{i \alpha B} +
  \eta_i^A \partial_{i B} - \tfrac{1}{4} \delta^A_B \theta_i^{\alpha
    C} \partial_{i \alpha C} - \tfrac{1}{4}\delta^A_B \eta_i^C \partial_{i C}
]\,,\notag\\
D &= \sum_i [-x_i^{\dot{\alpha}\alpha}\partial_{i \alpha \dot{\alpha}} -
  \tfrac{1}{2} \theta_i^{\alpha A} \partial_{i \alpha A} -
  \tfrac{1}{2} \lambda_i^{\alpha} \partial_{i \alpha} -\tfrac{1}{2}
  \tilde{\lambda}_i^{\dot{\alpha}} \partial_{i \dot{\alpha}}]\,,\notag\\
C &=  \sum_i [-\tfrac{1}{2}\lambda_i^{\alpha} \partial_{i \alpha} +
  \tfrac{1}{2}\tilde{\lambda}_i^{\dot{\alpha}} \partial_{i \dot{\alpha}} + \tfrac{1}{2}\eta_i^A
  \partial_{i A}]\,, \notag\\
S_{\alpha}^A &= \sum_i [-\theta_{i \alpha}^{B} \theta_i^{\beta A}
  \partial_{i \beta B} + x_{i \alpha}{}^{\dot{\beta}} \theta_i^{\beta
    A} \partial_{i \beta \dot{\beta}} + \lambda_{i \alpha}
  \theta_{i}^{\gamma A} \partial_{i \gamma} + x_{i+1\,
    \alpha}{}^{\dot{\beta}} \eta_i^A \partial_{i \dot{\beta}} -
  \theta_{i+1\, \alpha}^B \eta_i^A \partial_{i B}]\,,\notag\\
\overline{S}_{\dot{\alpha} A} &= \sum_i [x_{i \dot{\alpha}}{}^{\beta}
  \partial_{i \beta A} + \tilde{\lambda}_{i \dot{\alpha}}
  \partial_{iA}]\,,\notag\\
K_{\alpha \dot{\alpha}} &= \sum_i [x_{i \alpha}{}^{\dot{\beta}} x_{i
    \dot{\alpha}}{}^{\beta} \partial_{i \beta \dot{\beta}} + x_{i
    \dot{\alpha}}{}^{\beta} \theta_{i \alpha}^B \partial_{i \beta B} +
  x_{i \dot{\alpha}}{}^{\beta} \lambda_{i \alpha} \partial_{i \beta}
  + x_{i+1 \,\alpha}{}^{\dot{\beta}} \tilde{\lambda}_{i \dot{\alpha}}
  \partial_{i \dot{\beta}} + \tilde{\lambda}_{i \dot{\alpha}} \theta_{i+1\,
    \alpha}^B \partial_{i B}]\,.
\label{dualsc}
\end{align}
Note that if we restrict the dual generators $\bar{Q},\bar{S}$ to the on-shell superspace they
become identical to the conventional generators $\bar s, \bar q$.


\begin{thebibliography}{99}

  
\bibitem{Chari:1991xx}
V.~Chari and A.~Pressley,
\textit{``{Fundamental representations of Yangians and singularities of
  R-matrices}''},
\textsf{J~reigne~agnew.~Math.~417,~87~(1991)}.

\bibitem{Bernard:1992ya}
  D.~Bernard,
  \textit{``{An Introduction to Yangian Symmetries}}'',
  Int.\ J.\ Mod.\ Phys.\  B {\bf 7} (1993) 3517
  [arXiv:hep-th/9211133].

\bibitem{MacKay:2004tc}
  N.~J.~MacKay,
   \textit{``{Introduction to Yangian symmetry in integrable field theory}}'',
  Int.\ J.\ Mod.\ Phys.\  A {\bf 20} (2005) 7189
  [arXiv:hep-th/0409183].

\bibitem{Drinfeld:1985rx}
  V.~G.~Drinfeld,
  \textit{``{Hopf algebras and the quantum Yang-Baxter equation}}",
  Sov.\ Math.\ Dokl.\  {\bf 32} (1985) 254
  [Dokl.\ Akad.\ Nauk Ser.\ Fiz.\  {\bf 283} (1985) 1060].
  
\bibitem{Drinfeld:1986in}
  V.~G.~Drinfeld,
  \textit{``{Quantum groups}}",
  J.\ Sov.\ Math.\  {\bf 41} (1988) 898
  [Zap.\ Nauchn.\ Semin.\  {\bf 155} (1986) 18].
  
\bibitem{Drinfeld:1987sy}
  V.~G.~Drinfeld,
 \textit{``{A New realization of Yangians and quantized affine algebras}}'',
  Sov.\ Math.\ Dokl.\  {\bf 36} (1988) 212.

\bibitem{Drummond:2008cr}
  J.~M.~Drummond and J.~M.~Henn,
 \textit{``{All tree-level amplitudes in N=4 SYM}}",
  arXiv:0808.2475 [hep-th].

\bibitem{Witten:2003nn}
  E.~Witten,
 \textit{``{Perturbative gauge theory as a string theory in twistor space}}",
  Commun.\ Math.\ Phys.\  {\bf 252} (2004) 189
  [arXiv:hep-th/0312171].
  
\bibitem{Korchemsky:2009jv}
  G.~P.~Korchemsky and E.~Sokatchev,
 \textit{``{Twistor transform of all tree amplitudes in N=4 SYM theory}}",
  arXiv:0907.4107 [hep-th].
  
\bibitem{Bargheer:2009qu}
  T.~Bargheer, N.~Beisert, W.~Galleas, F.~Loebbert and T.~McLoughlin,
  \textit{``{Exacting N=4 Superconformal Symmetry}}'',
  JHEP {\bf 0911} (2009) 056
  [arXiv:0905.3738 [hep-th]].
  
\bibitem{Korchemsky:2009hm}
  G.~P.~Korchemsky and E.~Sokatchev,
 \textit{``{Symmetries and analytic properties of scattering amplitudes in N=4 SYM theory}}",
  arXiv:0906.1737 [hep-th].

\bibitem{Sever:2009aa}
  A.~Sever and P.~Vieira,
 \textit{``{Symmetries of the N=4 SYM S-matrix}}",
  arXiv:0908.2437 [hep-th].

\bibitem{Beisert:2010gn}
  N.~Beisert, J.~Henn, T.~McLoughlin and J.~Plefka,
 \textit{``{One-Loop Superconformal and Yangian Symmetries of Scattering Amplitudes in N=4 Super Yang-Mills}}",
  arXiv:1002.1733 [hep-th].
  

\bibitem{Drummond:2010qh}
  J.~M.~Drummond and L.~Ferro,
  \textit{``{Yangians, Grassmannians and T-duality}}'',
  JHEP {\bf 1007} (2010) 027
  [arXiv:1001.3348 [hep-th]].

  
\bibitem{Drummond:2008vq}
  J.~M.~Drummond, J.~Henn, G.~P.~Korchemsky and E.~Sokatchev,
 \textit{``{Dual superconformal symmetry of scattering amplitudes in N=4 super-Yang-Mills theory}}",
  arXiv:0807.1095 [hep-th].

\bibitem{Berkovits:2008ic}
  N.~Berkovits and J.~Maldacena,
 \textit{``{Fermionic T-Duality, Dual Superconformal Symmetry, and the Amplitude/Wilson Loop Connection}}",
  JHEP {\bf 0809} (2008) 062
  [arXiv:0807.3196 [hep-th]].

\bibitem{Beisert:2008iq}
  N.~Beisert, R.~Ricci, A.~A.~Tseytlin and M.~Wolf,
 \textit{``{Dual Superconformal Symmetry from AdS5 x S5 Superstring Integrability}}",
  Phys.\ Rev.\  D {\bf 78} (2008) 126004
  [arXiv:0807.3228 [hep-th]].


\bibitem{Beisert:2009cs}
  N.~Beisert,
 \textit{``{T-Duality, Dual Conformal Symmetry and Integrability for Strings on $AdS_5 \mathrm{x} S^5$}}",
  Fortsch.\ Phys.\  {\bf 57} (2009) 329
  [arXiv:0903.0609 [hep-th]].

\bibitem{Brandhuber:2008pf}
  A.~Brandhuber, P.~Heslop and G.~Travaglini,
 \textit{``{A note on dual superconformal symmetry of the N=4 super Yang-Mills S-matrix}}",
  Phys.\ Rev.\  D {\bf 78} (2008) 125005
  [arXiv:0807.4097 [hep-th]].

\bibitem{Drummond:2009fd}
  J.~M.~Drummond, J.~M.~Henn and J.~Plefka,
 \textit{``{Yangian symmetry of scattering amplitudes in N=4 super Yang-Mills theory}}",
  JHEP {\bf 0905} (2009) 046
  [arXiv:0902.2987 [hep-th]].

\bibitem{AdS/INT1}
  J.~A.~Minahan and K.~Zarembo,
  \textit{``{The Bethe-ansatz for ${\cal N}=4$ super Yang-Mills}}'',
  JHEP {\bf 0303}, 013 (2003)
  [hep-th/0212208].

\bibitem{AdS/INT1b}
  N.~Beisert, C.~Kristjansen and M.~Staudacher,
  \textit{``{The dilatation operator of ${\cal N}=4$ super Yang-Mills theory}}'',
  Nucl.\ Phys.\  B {\bf 664}, 131 (2003)
  [hep-th/0303060].

\bibitem{Beisert:2005fw}
  N.~Beisert and M.~Staudacher,
  \textit{``{Long-range PSU(2,2|4) Bethe ansaetze for gauge theory and strings''}},
  Nucl.\ Phys.\  B {\bf 727} (2005) 1
  [arXiv:hep-th/0504190].



\bibitem{Bena:2003wd}
  I.~Bena, J.~Polchinski and R.~Roiban,
  \textit{``{Hidden symmetries of the AdS(5) x S**5 superstring}}'',
  Phys.\ Rev.\  D {\bf 69} (2004) 046002
  [arXiv:hep-th/0305116].

\bibitem{Dolan:2004ps}
  L.~Dolan, C.~R.~Nappi and E.~Witten,
 \textit{``{Yangian symmetry in D = 4 superconformal Yang-Mills theory}}",
  arXiv:hep-th/0401243.
 

\bibitem{Hodges:2009hk}
  A.~Hodges,
 \textit{``{Eliminating spurious poles from gauge-theoretic amplitudes}}",
  arXiv:0905.1473 [hep-th].

\bibitem{ArkaniHamed:2009dn}
  N.~Arkani-Hamed, F.~Cachazo, C.~Cheung and J.~Kaplan,
 \textit{``{A Duality For The S Matrix}}",
  arXiv:0907.5418 [hep-th].

\bibitem{Mason:2009qx}
  L.~Mason and D.~Skinner,
 \textit{``{Dual Superconformal Invariance, Momentum Twistors and Grassmannians}}",
  arXiv:0909.0250 [hep-th].


\bibitem{Alday:2009zm}
  L.~F.~Alday, J.~M.~Henn, J.~Plefka and T.~Schuster,
 \textit{``{Scattering into the fifth dimension of N=4 super Yang-Mills}}",
  arXiv:0908.0684 [hep-th].

\bibitem{Henn:2010bk}
  J.~M.~Henn, S.~G.~Naculich, H.~J.~Schnitzer and M.~Spradlin,
   \textit{``{Higgs-regularized three-loop four-gluon amplitude in N=4 SYM: exponentiation and Regge limits}}'',
  JHEP {\bf 1004} (2010) 038
  [arXiv:1001.1358 [hep-th]].

\bibitem{Henn:2010ir}
  J.~M.~Henn, S.~G.~Naculich, H.~J.~Schnitzer and M.~Spradlin,
   \textit{``{More loops and legs in Higgs-regularized N=4 SYM amplitudes}}'',
  JHEP {\bf 1008} (2010) 002
  [arXiv:1004.5381 [hep-th]].

\bibitem{Bern:2010qa}
  Z.~Bern, J.~J.~Carrasco, T.~Dennen, Y.~t.~Huang and H.~Ita,
    \textit{``{Generalized Unitarity and Six-Dimensional Helicity}}'',
  Phys.\ Rev.\  D {\bf 83} (2011) 085022
  [arXiv:1010.0494 [hep-th]].

\bibitem{CaronHuot:2010rj}
  S.~Caron-Huot and D.~O'Connell,
    \textit{``{Spinor Helicity and Dual Conformal Symmetry in Ten Dimensions}}'',
  arXiv:1010.5487 [hep-th].
  
\bibitem{Dennen:2010dh}
  T.~Dennen and Y.~t.~Huang,
    \textit{``{Dual Conformal Properties of Six-Dimensional Maximal Super Yang-Mills Amplitudes}}'',
  JHEP {\bf 1101} (2011) 140
  [arXiv:1010.5874 [hep-th]].

\bibitem{Beisert:2011pn}
  N.~Beisert and B.~U.~W.~Schwab,
   \textit{``{Bonus Yangian Symmetry for the Planar S-Matrix of N=4 Super Yang-Mills}}'',
  arXiv:1103.0646 [hep-th].
  

\bibitem{Belitsky:2011zm}
  A.~V.~Belitsky, G.~P.~Korchemsky and E.~Sokatchev,
  \textit{``{Are scattering amplitudes dual to super Wilson loops?}}'',
  arXiv:1103.3008 [hep-th].



\bibitem{Mason:2010yk}
  L.~J.~Mason and D.~Skinner,
  \textit{``{The Complete Planar S-matrix of N=4 SYM as a Wilson Loop in Twistor Space}}'',
  JHEP {\bf 1012} (2010) 018
  [arXiv:1009.2225 [hep-th]].

\bibitem{CaronHuot:2010ek}
  S.~Caron-Huot,
    \textit{``{Notes on the scattering amplitude / Wilson loop duality}}'',
  arXiv:1010.1167 [hep-th].

\bibitem{Ivanov:1985bk}
  S.~V.~Ivanov and G.~P.~Korchemsky,
   \textit{``{Some Supplements of Nonperturbative Gauges}}'',
  Phys.\ Lett.\  B {\bf 154} (1985) 197.

\bibitem{Ivanov:1985np}
  S.~V.~Ivanov, G.~P.~Korchemsky and A.~V.~Radyushkin,
   \textit{``{Infrared Asymptotics Of Perturbative Qcd: Contour Gauges}}'',
  Yad.\ Fiz.\  {\bf 44}, 230 (1986)
  [Sov.\ J.\ Nucl.\ Phys.\  {\bf 44}, 145 (1986)].

\bibitem{Korchemsky:1985xj}
  G.~P.~Korchemsky and A.~V.~Radyushkin,
  \textit{``{Loop Space Formalism And Renormalization Group For The Infrared Asymptotics Of Qcd}}'',
  Phys.\ Lett.\  B {\bf 171}, 459 (1986).

\bibitem{Bern:2005iz}
  Z.~Bern, L.~J.~Dixon and V.~A.~Smirnov,
    \textit{``{Iteration of planar amplitudes in maximally supersymmetric Yang-Mills theory at three loops and beyond}}'',
  Phys.\ Rev.\  D {\bf 72}, 085001 (2005)
  [arXiv:hep-th/0505205].


\bibitem{Alday:2007hr}
  L.~F.~Alday and J.~M.~Maldacena,
  \textit{``{Gluon scattering amplitudes at strong coupling}}'',
  JHEP {\bf 0706} (2007) 064
  [arXiv:0705.0303 [hep-th]].


\bibitem{Drummond:2007aua}
  J.~M.~Drummond, G.~P.~Korchemsky and E.~Sokatchev,
  \textit{``{Conformal properties of four-gluon planar amplitudes and Wilson loops}}",
  Nucl.\ Phys.\  B {\bf 795} (2008) 385
  [arXiv:0707.0243 [hep-th]].

\bibitem{Brandhuber:2007yx}
  A.~Brandhuber, P.~Heslop and G.~Travaglini,
  \textit{``{MHV Amplitudes in N=4 Super Yang-Mills and Wilson Loops}}'',
  Nucl.\ Phys.\  B {\bf 794} (2008) 231
  [arXiv:0707.1153 [hep-th]].

\bibitem{Drummond:2007cf}
  J.~M.~Drummond, J.~Henn, G.~P.~Korchemsky and E.~Sokatchev,
  \textit{``{On planar gluon amplitudes/Wilson loops duality}}",
  Nucl.\ Phys.\  B {\bf 795} (2008) 52
  [arXiv:0709.2368 [hep-th]].


\bibitem{Drummond:2007au}
  J.~M.~Drummond, J.~Henn, G.~P.~Korchemsky and E.~Sokatchev,
  \textit{``{Conformal Ward identities for Wilson loops and a test of the duality with gluon amplitudes}}'',
  Nucl.\ Phys.\  B {\bf 826}, 337 (2010)
  [arXiv:0712.1223 [hep-th]].

\bibitem{Drummond:2007bm}
  J.~M.~Drummond, J.~Henn, G.~P.~Korchemsky and E.~Sokatchev,
  \textit{``{The hexagon Wilson loop and the BDS ansatz for the six-gluon amplitude'}}",
  Phys.\ Lett.\  B {\bf 662} (2008) 456
  [arXiv:0712.4138 [hep-th]].

\bibitem{Drummond:2008aq}
  J.~M.~Drummond, J.~Henn, G.~P.~Korchemsky and E.~Sokatchev,
  \textit{``{Hexagon Wilson loop = six-gluon MHV amplitude}}'',
  Nucl.\ Phys.\  B {\bf 815} (2009) 142
  [arXiv:0803.1466 [hep-th]].


\bibitem{Bern:2008ap}
  Z.~Bern, L.~J.~Dixon, D.~A.~Kosower, R.~Roiban, M.~Spradlin, C.~Vergu and A.~Volovich,
   \textit{``{The Two-Loop Six-Gluon MHV Amplitude in Maximally Supersymmetric Yang-Mills Theory}}'',
  Phys.\ Rev.\  D {\bf 78}, 045007 (2008)
  [arXiv:0803.1465 [hep-th]].


\bibitem{DelDuca:2010zg}
  V.~Del Duca, C.~Duhr and V.~A.~Smirnov,
  \textit{``{The Two-Loop Hexagon Wilson Loop in N = 4 SYM}}'',
  JHEP {\bf 1005}, 084 (2010)
  [arXiv:1003.1702 [hep-th]].

\bibitem{DelDuca:2010zp}
  V.~Del Duca, C.~Duhr and V.~A.~Smirnov,
\textit{``{A Two-Loop Octagon Wilson Loop in N = 4 SYM}}'',
  JHEP {\bf 1009} (2010) 015
  [arXiv:1006.4127 [hep-th]].



\bibitem{Goncharov:2010jf}
  A.~B.~Goncharov, M.~Spradlin, C.~Vergu and A.~Volovich,
  \textit{``{Classical Polylogarithms for Amplitudes and Wilson Loops}}",
  Phys.\ Rev.\ Lett.\  {\bf 105}, 151605 (2010)
  [arXiv:1006.5703 [hep-th]].

\bibitem{CaronHuot:2011ky}
  S.~Caron-Huot,
  \textit{``{Superconformal symmetry and two-loop amplitudes in planar N=4 super Yang-Mills}}",
  arXiv:1105.5606 [hep-th].


\bibitem{ArkaniHamed:2009vw}
  N.~Arkani-Hamed, F.~Cachazo and C.~Cheung,
 \textit{``{The Grassmannian Origin Of Dual Superconformal Invariance}}",
  arXiv:0909.0483 [hep-th].


\bibitem{Drummond:2010uq}
  J.~M.~Drummond and L.~Ferro,
  \textit{``{The Yangian origin of the Grassmannian integral}}'',
  JHEP {\bf 1012} (2010) 010
  [arXiv:1002.4622 [hep-th]].

\bibitem{Korchemsky:2010ut}
  G.~P.~Korchemsky and E.~Sokatchev,
  \textit{``{Superconformal invariants for scattering amplitudes in N=4 SYM theory}}'',
  Nucl.\ Phys.\  B {\bf 839} (2010) 377
  [arXiv:1002.4625 [hep-th]].

\bibitem{ArkaniHamed:2010kv}
  N.~Arkani-Hamed, J.~L.~Bourjaily, F.~Cachazo, S.~Caron-Huot and J.~Trnka,
  \textit{``{The All-Loop Integrand For Scattering Amplitudes in Planar N=4 SYM}}'',
  JHEP {\bf 1101} (2011) 041
  [arXiv:1008.2958 [hep-th]].

  
\bibitem{Drummond:2010zv}
  J.~M.~Drummond, L.~Ferro and E.~Ragoucy,
  \textit{``{Yangian symmetry of light-like Wilson loops}}",
  arXiv:1011.4264 [hep-th].

\bibitem{Alday:2009yn}
  L.~F.~Alday and J.~Maldacena,
  \textit{``{Null polygonal Wilson loops and minimal surfaces in Anti-de-Sitter space}}'',
  JHEP {\bf 0911}, 082 (2009)
  [arXiv:0904.0663 [hep-th]].

\bibitem{Alday:2010ku}
  L.~F.~Alday, D.~Gaiotto, J.~Maldacena, A.~Sever and P.~Vieira,
  \textit{``{An Operator Product Expansion for Polygonal null Wilson Loops}}'',
  JHEP {\bf 1104}, 088 (2011)
  [arXiv:1006.2788 [hep-th]].

\bibitem{Bern:1994zx}
  Z.~Bern, L.~J.~Dixon, D.~C.~Dunbar and D.~A.~Kosower,
  \textit{``{One-Loop n-Point Gauge Theory Amplitudes, Unitarity and Collinear Limits}}",
  Nucl.\ Phys.\  B {\bf 425} (1994) 217
  [arXiv:hep-ph/9403226].


\bibitem{Heslop:2010kq}
  P.~Heslop and V.~V.~Khoze,
  \textit{``{Analytic Results for MHV Wilson Loops}}'',
  JHEP {\bf 1011}, 035 (2010)
  [arXiv:1007.1805 [hep-th]].

\bibitem{Gaiotto:2010fk}
  D.~Gaiotto, J.~Maldacena, A.~Sever and P.~Vieira,
  \textit{``{Bootstrapping Null Polygon Wilson Loops}}'',
  JHEP {\bf 1103}, 092 (2011)
  [arXiv:1010.5009 [hep-th]].



\end{thebibliography}
\end{document}